\documentclass[11pt]{elsarticle}
\usepackage[margin=1.2in]{geometry}
\usepackage{gensymb}
\usepackage{graphicx}
\usepackage{amsmath}
\usepackage{latexsym}
\usepackage{amsfonts}
\usepackage{mathtools}
\usepackage{tcolorbox}
\usepackage{listings}
\usepackage{hyperref}
\hypersetup{
    colorlinks=true,
    linkcolor=blue,
    filecolor=magenta,      
    urlcolor=cyan,
    pdftitle={Overleaf Example},
    pdfpagemode=FullScreen,
    }
\usepackage{array}
\usepackage{subcaption}
\usepackage{kantlipsum}
\usepackage{changepage}
\usepackage{dirtytalk}
\usepackage{natbib}
\usepackage[makeroom]{cancel}
\usepackage{amsthm}
\theoremstyle{remark}

\usepackage{gensymb}

\usepackage{relsize}
\usepackage{float}
\usepackage{pifont}
\usepackage[font={small}]{caption}
\DeclareCaptionFormat{myformat}{\fontsize{9}{10}\selectfont#1#2#3}
\makeatletter
\renewcommand{\fnum@figure}{\textbf{Fig. \thefigure.}}
\makeatother
\makeatletter
\renewcommand{\fnum@table}{\textbf{Tab. \thetable.}}
\makeatother
\usepackage[labelsep=space]{caption}

\usepackage{amssymb}
\usepackage{a4wide}
\usepackage{pst-all}
\usepackage{float}
\usepackage{latexsym}
\usepackage{graphicx}
\usepackage{psfrag}
\usepackage{verbatim}
\usepackage[ruled,vlined]{algorithm2e}
\usepackage{setspace}
\usepackage{xcolor,colortbl}

\newcommand{\eqrefK}[1]{\mbox{Eq. (\ref{#1})}}
\newcommand{\figref}[1]{\mbox{Fig. \ref{#1}}}

\renewcommand*\thetable{\Roman{table}}

\newfloat{MyBox}{thp}{loa}
\floatname{MyBox}{Box}

\usepackage{upgreek}

\makeatletter
\def\@author#1{\g@addto@macro\elsauthors{\normalsize%
    \def\baselinestretch{1}%
    \upshape\authorsep#1\unskip\textsuperscript{%
      \ifx\@fnmark\@empty\else\unskip\sep\@fnmark\let\sep=,\fi
      \ifx\@corref\@empty\else\unskip\sep\@corref\let\sep=,\fi
      }%
    \def\authorsep{\unskip,\space}%
    \global\let\@fnmark\@empty
    \global\let\@corref\@empty  
    \global\let\sep\@empty}%
    \@eadauthor={#1}
}
\makeatother

\newcommand{\ds}{\displaystyle}
\newcommand{\bm}[1]{\text{\boldmath $#1$}}

\newcommand{\norm}[1]{\left\lVert#1\right\rVert}
\usepackage{tikz}

\newcommand{\tripledot}{%
\tikz[baseline=-0.2ex]{ \draw[black,fill=black] (0,0) circle (.1ex); \draw[black,fill=black] (0,0.6ex) circle (.1ex); \draw[black,fill=black] (0,1.2ex) circle (.1ex) }%
}

\makeatletter
\def\ps@pprintTitle{%
 \let\@oddhead\@empty
 \let\@evenhead\@empty
 \def\@oddfoot{\centerline{\thepage}}
 \let\@evenfoot\@oddfoot}
\makeatother

\makeatletter
\def\ps@pprintTitle{%
 \let\@oddhead\@empty
 \let\@evenhead\@empty
 \def\@oddfoot{}%
 \let\@evenfoot\@oddfoot}
\makeatother

\begin{document}

\begin{frontmatter}
\title{Topology optimization of contact-aided thermo-mechanical regulators}
\author{Anna Dalklint\fnref{label1}\corref{cor1}}
\author{Joe Alexandersen \fnref{label2}}
\author{Andreas Henrik Frederiksen\fnref{label1}}
\author{Konstantinos Poulios\fnref{label1}}
\author{Ole Sigmund\fnref{label1}}

\address[label1]{Technical University of Denmark, Department of Civil and Mechanical Engineering, Nils Koppels Allé, Building 404, 2800 Kongens
Lyngby, Denmark \fnref{label1}}
\address[label2]{University of Southern Denmark, Department of Mechanical and Electrical Engineering, Odense 5230, Denmark \fnref{label2}}
\cortext[cor1]{Corresponding author. E-mail adress: adal@dtu.dk}

\begin{abstract}
Topology optimization is used to systematically design contact-aided thermo-mechanical regulators, i.e. components whose effective thermal conductivity is tunable by mechanical deformation and contact. The thermo-mechanical interactions are modeled using a fully coupled non-linear thermo-mechanical finite element framework. To obtain the intricate heat transfer response, the components leverage self-contact, which is modeled using a third medium contact method. The effective heat transfer properties of the regulators are tuned by solving a topology optimization problem using a traditional gradient based algorithm. Several designs of thermo-mechanical regulators in the form of switches, diodes and triodes are presented.

\end{abstract}

\begin{keyword}
Thermo-mechanical regulators \sep Thermal components \sep Coupled non-linear thermo-mechanical analysis \sep Third medium contact
\end{keyword}

\end{frontmatter}


\section{Introduction}
Managing heat transfer poses a significant challenge in modern technology.
While this challenge always has existed for conventional heating and cooling devices like heat exchangers and refrigerators, it has become increasingly important in modern applications to e.g. enhance the lifespan of electronics, improve advanced manufacturing processes, minimize energy consumption in high-performance computing or data storage clusters, and enable certain functionalities in space applications. Traditionally, the thermal properties of these systems are regulated using linear and passive thermal components, whose thermal properties are mainly governed by their static thermal conductivity and size. However, there is a growing interest in exploring components that harness switchable and non-linear heat transfer mechanisms to obtain tunable heat transfer control. This development has lead to the realization of advanced components such as thermal switches, diode and triodes, cf. \citet{wehmeyer2017thermal} and \citet{castelli2023three}. In this work, we refer to components that exhibit non-linear heat transfer functions as \emph{thermal regulators}.

Thermal regulators are useful in various applications. For example, they can be used to actively regulate the indoor climate of buildings (\citet{miao2022non}), or actively manage heat dissipation rates in e.g. electronics (\citet{yang2019integrated}) and spacecrafts (\citet{milanez2003theoretical}). See review articles by e.g. \citet{wehmeyer2017thermal} as well as \citet{wong2021review} and \citet{ram2023critical} for an exhaustive exploration of application possibilities. Actuation mechanisms and thereby design of the thermal regulators can vary. One approach to the design of thermal regulators relies on active actuation, i.e. when the effective conductivity of the regulator is actively controlled via e.g. applied pressure (\citet{cho2007fabrication}), electric fields (\citet{almanza2018electrostatically}) or magnetic fields (\citet{bosisio2015magnetic}). Another approach relies on passive actuation, such that the component self-regulates in response to the exterior temperature changes via thermal expansion/shrinking (\citet{heo2019passive}) or material phase changes (\citet{gulfam2019advanced}). Existing designs of regulators often leverage conduction via moving structural parts, i.e. mechanics, (\citet{cho2007fabrication}), or gas/fluid convection (\citet{yang2019integrated}) to change their effective thermal conductivity during actuation. In this work, we focus on the former, i.e. \emph{thermo-mechanical} regulators.

\begin{figure}[H]
\centering
\includegraphics[width=0.45\textwidth]{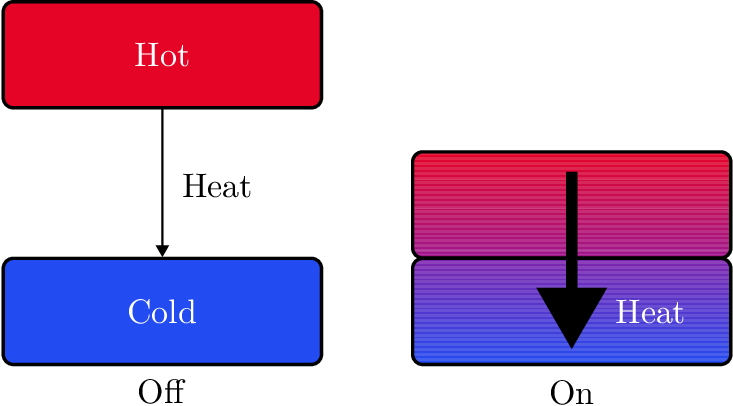}
\caption{A schematic of a thermal switch leveraging contact.}
\label{fig:switchSchematic}
\end{figure}

A thermal regulator should ultimately exhibit a non-linear or switchable heat transfer function, but the desired shape of this transfer function depends on the application. For example, a \emph{thermal switch} exhibits two distinct functional states: an \say{on} state facilitating heat transfer and an \say{off} state where heat transfer is impeded. A \emph{thermal diode}, on the other hand, should rectify the heat flow, such that the heat transfer along a specific axis is facilitated or impeded depending on the sign of the temperature gradient. An even more complex component is the \emph{thermal triode}, which, analogously to an electrical triode/transistor, should be able to amplify an incoming signal. A natural and efficient way of controlling the effective thermal conductivity in a thermo-mechanical regulator is to develop designs that utilize structural contact. In this way, the effective heat conductivity is high when contact occurs, and otherwise low, cf. the schematic of a contact-aided thermal switch in \figref{fig:switchSchematic}.
\noindent

In the existing literature, thermal regulators are designed based on human intuition and time-consuming design-and-experimental iterations. This process can prove cumbersome when designing regulators with highly complex heat transfer functions. Fortunately, this design process is an obvious candidate for more efficient systematic design approaches such as topology optimization. One hurdle to the topology optimization of contact-aided thermal regulators is however their modeling complexity, caused by 1) the coupled thermo-mechanics and 2) the intrinsically non-linear contact physics. 

The topology optimization of thermo-mechanical problems appears in numerous works. For example, \citet{sigmund2001design1, sigmund2001design2} designed thermal actuators undergoing finite deformations using topology optimization, which later were realized by \citet{jonsmann1999compliant}. Recently, \citet{granlund2024topology} considered the topology optimization of coupled transient non-linear thermo-mechanical problems. A topology optimization approach to inverse motion based form finding of coupled static thermo-mechanical problems was investigated by \citet{sui2023topology}. \citet{chung2020level} used a level-set approach to the topology optimization of structures undergoing large deformations due to thermal and mechanical loads.

Recently, \citet{bluhm2021internal} illustrated the power of the so-called third medium contact method in topology optimization problems of contact-aided mechanisms. This contact model, pioneered by \citet{wriggers2013finite}, approximates contact pressures via the reactive tractions that occur when a third medium, that must be present in between the contacting surfaces, is compressed. Indeed, via appropriate constitutive modeling, the third medium naturally exists in the topology optimization formulation in the form of the ersatz material. \citet{bluhm2021internal} employed a monolithic approach to the topology optimization problem, i.e. solved the minimization of the objective and physical residuals simultaneously using Newton's method. In a recent work, \citet{Andreas2023topology} instead implemented the approach in the traditional staggered fashion, and use MMA (Method of Moving Asymptotes, cf. \citet{svanberg1987method}) to update the design. Other examples of the the third medium contact method in topology optimization include \citet{bluhm2023inverse}, whereby non-linear springs are designed, and \citet{dalklint2023computational}, who design metamaterials that utilized self-contact to obtain a prescribed macroscopic stress-strain behavior. Very recently, \citet{faltus2024third} extended a third medium contact model to pneumatically actuated systems.

To summarize; in this work, we use topology optimization to design thermo-mechanical regulators that harness self-contact to exhibit highly non-linear heat transfer functions. To this end, we extend the prospect of the third medium contact method to thermo-mechanical problems, such that we now seek a third medium that, in addition to contact tractions, transfers heat when contact occurs, but otherwise approximates an isolating material. The finite element method is used to model the fully coupled non-linear thermo-mechanical problem. The optimization problems are solved using the gradient-based optimizer MMA, and the gradients of the cost and constraint functions are obtained from an adjoint sensitivity analysis. The numerical results display various designs of thermo-mechanical switches, diodes and triodes.

\section{Preliminaries}
In our steady-state thermo-mechanical formulation, the reference, undeformed configuration $\Omega\in\mathbb{R}^2$ and the current, deformed configuration $\Omega_c\in\mathbb{R}^2$ are distinguished. A material point $\bm{X}\in\Omega$ is displaced to its deformed, spatial position $\bm{x}\in\Omega_c$ via the displacement field $\bm{u}(\bm{X}):\Omega\rightarrow \mathbb{R}^2$, such that $\bm{x} = \bm{X} +\bm{u}(\bm{X})$. The local deformation is described by the deformation gradient $\bm{F}=\bm{\nabla}\bm{x} = \bm{1}+\bm{\nabla}\bm{u}$, where $\bm{\nabla}$ is the gradient operator with respect to $\bm{X}$, and $J=\text{det}(\bm{F})$. The temperature field is denoted $\theta(\bm{X}):\Omega\rightarrow\mathbb{R}$. The deformation gradient is assumed to satisfy the multiplicative split $\bm{F}=\bm{F}^e\bm{F}^\theta$, where $\bm{F}^e$ and $\bm{F}^\theta$ are the elastic and thermal parts of the deformation gradient, respectively. An isotropic thermal deformation is assumed, i.e. $\bm{F}^\theta = \left(J^\theta\right)^{1/3}\bm{1}$ with $J^\theta = \text{det}(\bm{F}^\theta) = e^{3\alpha(\theta-\theta_o)}$, where $\alpha$ is the thermal expansion coefficient and $\theta_o$ is the reference temperature. For future reference, we denote $J^e = \text{det}(\bm{F}^e)$.
 
The undeformed boundary $\partial\Omega$ with normal $\bm{n}$, is decomposed into the complementary surfaces $\partial\Omega_u$ and $\partial\Omega_t$ for the mechanical and $\partial\Omega_\theta$, $\partial\Omega_{q}$ and $\partial\Omega_{q_c}$ for the thermal boundary value problems. For the mechanical problem, we prescribe the displacement $\bm{u}=\bar{\bm{u}}$ on $\partial\Omega_{u}$ and the traction $\bm{t}=\bm{0}$ on $\partial\Omega_t$, whereas for the thermal problem we prescribe the temperature $\theta=\bar{\theta}$ on $\partial\Omega_{\theta}$, the heat flux $\bm{q}\cdot\bm{n} = 0$ on $\partial\Omega_{q}$ and Newton convection $\bm{q}\cdot\bm{n} = h\left(\theta -\theta_\infty \right)$ on $\partial\Omega_{q_c}$, where $h$ is the heat transfer coefficient and $\theta_\infty$ is the temperature of the surrounding medium. For simplicity, it is always assumed that $\bm{u}=\bm{0}$ on $\partial\Omega_{q_c}$. The heat flux in the undeformed configuration follows from Fourier's law, i.e. 
\begin{equation}
\bm{q} = -J\bm{F}^{-1}\bm{\kappa}\bm{F}^{-T}\cdot \bm{\nabla}\theta,
\label{heatFlux}
\end{equation}
where $\bm{\kappa} = \kappa\bm{1}$ is the spatial isotropic heat conductivity tensor and $\bm{1}$ is the identity. 

Following \citet{miehe1995entropic} and \citet{holzapfel2000nonlinear}, we assume reversible, elastic, mechanical response, such that the weak forms corresponding to the thermo-mechanical problem are
\begin{equation}
\begin{array}{ll}
\ds\mathcal{R}_u(\bm{u},\theta,\bm{\lambda}_u;\delta\bm{u}) &\ds=  \int_\Omega \Psi_{,u}(\bm{u},\theta;\delta \bm{u}) \, dV + \int_{\partial\Omega_{u}}\bm{\lambda}_u\cdot\delta\bm{u} \, dS= 0, \\[10pt]
\ds\mathcal{R}_\theta(\bm{u},\theta, \lambda_\theta;\delta\theta) &\ds= 
\int_\Omega \bm{\nabla}\delta\theta\cdot\bm{q}(\bm{u},\theta) \, dV +
\int_{\partial\Omega_{\theta}}\lambda_\theta \cdot\delta\theta \, dS
+ \int_{\partial\Omega_{q_c}}h\left(\theta-\theta_\infty\right)\delta\theta\, dS = 0, \\[10pt]
\ds\mathcal{R}_{\lambda_u}(\bm{u}, \bm{\lambda}_u;\delta\bm{\lambda}_u) &\ds=  \int_{\partial\Omega_{u}} \left(\bm{u}-\bar{\bm{u}}\right)\cdot\delta\bm{\lambda}_u \, dS = 0, \\[10pt]
\ds\mathcal{R}_{\lambda_\theta}(\theta, \lambda_\theta;\delta\lambda_\theta) &\ds=
\int_{\partial\Omega_{\theta}} \left(\theta-(\theta_o+\bar{\theta})\right )\cdot\delta\lambda_\theta\, dS = 0,
\end{array}
\label{weak1}
\end{equation}
for all $(\delta\bm{u},\delta\theta,\delta\bm{\lambda}_u,\delta\lambda_\theta)\in(\mathbb{R}^2\times\mathbb{R}\times\mathbb{R}^2\times\mathbb{R})$, and where the boundary conditions are applied in a weak sense via the Lagrange multiplier fields $\bm{\lambda}_u\in\partial\Omega_{u}$ and $\lambda_\theta\in\partial\Omega_{\theta}$. It is noted that $\bm{\lambda}_u$ and $\lambda_\theta$ represents unknown reaction tractions and normal heat fluxes, i.e. $\bm{n}\cdot\bm{q}$, respectively. In the above, the Kelvin-Joule effect is neglected and vanishing heat sources are assumed within $\Omega$. We further let $\Psi$ denote the strain energy density, which is choosen according to the isotropic neo-Hookean type hyperelastic model
\begin{equation}
\begin{array}{ll}
\ds\Psi(\bm{u},\theta) &\ds= \frac{K}{2}\text{ln}(J^e)^2 + \frac{G}{2}\left((J^e)^{-2/3}\norm{\bm{F}^e}^2 -3 \right)  \\[10pt]
&\ds= \frac{K}{2}\left(\text{ln}(J)^2 - 6\alpha(\theta-\theta_o)\text{ln}(J)  + 9\alpha^2(\theta-\theta_o)^2 \right) + \frac{G}{2}\left(J^{-2/3}\norm{\bm{F}}^2 -3 \right), 
\end{array}
\label{Psi}
\end{equation}
where $K=E/3(1-2\nu)$, $G=E/2(1+\nu)$, $E$ and $\nu$ denote the initial bulk modulus, shear modulus, Young's modulus and Poisson's ratio in the limit of infinitesimal strains. From the strain energy density in \eqrefK{Psi}, the first Piola-Kirchhoff stress tensor is computed as
\begin{equation}
\bm{P}(\bm{\nabla}\bm{u}) = \frac{\partial\Psi}{\partial \bm{F}} = K \left(\text{ln}(J) \bm{F}^{-T} - 3\alpha(\theta-\theta_o) \bm{F}^{-T} \right)+ G J^{-2/3}\text{dev}(\bm{F}\bm{F}^T)\bm{F}^{-T},
\label{PK1}
\end{equation}
where $\text{dev}(\bm{A}) = \bm{A} - \frac{1}{3}\text{tr}(\bm{A})\bm{1}$ for an arbitrary second order tensor $\bm{A}$. In this way, we find $\Psi_{,u} = \bm{P} : \bm{\nabla}\delta\bm{u}$, cf. \eqrefK{weak1}. For later convenience, we define 
\begin{equation}
\mathcal{R}(\bm{a};\delta\bm{a})  = 
\mathcal{R}_u(\bm{u},\theta,\bm{\lambda}_u;\delta\bm{u}) +
\mathcal{R}_\theta(\bm{u},\theta, \lambda_\theta;\delta\theta) +
\mathcal{R}_{\lambda_u}(\bm{u}, \bm{\lambda}_u;\delta\bm{\lambda}_u) +
\mathcal{R}_{\lambda_\theta}(\theta, \lambda_\theta;\delta\lambda_\theta)
= 0,
\label{allR}
\end{equation}
where $\bm{a} = (\bm{u},\theta,\bm{\lambda}_u,\lambda_\theta)$.

\section{Model components}
\subsection{Topology optimization}
The goal of the topology optimization is to distribute the material in the design domain $\Omega$, such that an objective function is minimized. To this end, let $z\in\Omega\rightarrow[0,1]$ denote the design variable volume fraction field. To regularize the optimization problem, the design space is restricted via the PDE filter (\citet{lazarov2011filters}), such that fine scale oscillations of $z$ are penalized. This provides a smoothed field $\zeta\in\Omega\rightarrow[0,1]$, as a solution to the following boundary value problem
\begin{equation}
\mathcal{R}_\zeta(\zeta,z;\delta\zeta) = \int_{\Omega} l^2\bm{\nabla}\zeta\cdot\bm{\nabla}\delta\zeta \, dV + \int_{\Omega} \left(\zeta-z\right)\delta\zeta \, dV = 0,
\label{Helmholtz}
\end{equation}
for all $\delta\zeta\in\mathbb{R}$. In the above, $l$ is the filter radius.

This filtering invariably produces regions wherein $\zeta(\bm{X})\in(0,1)$. In order to push the solution towards a \say{black and white} design, thresholding (\citet{guest2004achieving}) is employed via the approximate Heaviside projection (\citet{wang2011projection})
\begin{equation}
\bar{\zeta} = H_{\beta,\eta}(\zeta) =  \frac{\text{tanh}(\beta\eta)+\text{tanh}(\beta(\zeta-\eta))}{\text{tanh}(\beta\eta)+\text{tanh}(\beta(1-\eta))},
\label{Heavi}
\end{equation}
where $\beta$ and $\eta=0.5$ are numerical parameters that control the slope and position of the Heaviside function. Additionally, intermediate $\zeta(\bm{X})\in(0,1)$ regions are implicitly penalized via 
the SIMP scheme (\citet{bendsoe1989optimal}), such that the material parameters ultimately are interpolated as
\begin{equation}
\begin{array}{cc}
\ds K = \delta_o^E + (1-\delta_o^E)\bar{\zeta}^p K_o, \ \ G =\delta_o^E + (1-\delta_o^E)\bar{\zeta}^pG_o, \\[10pt] 
\kappa =\delta_o^\kappa + (1-\delta_o^\kappa)\bar{\zeta}^p\kappa_o \ \ \text{and} \ \ K\alpha  =\bar{\zeta}^pK\alpha_o. 
\end{array}
\label{matInt}
\end{equation}
In the above, $K_o,G_o,\kappa_o$ and $\alpha_o$ are the reference bulk-, shear modulus, thermal conductivity and thermal expansion coefficient, respectively. Also, $p \geq 0$ is the penalization parameter and $\delta_o^E, \delta_o^\kappa \geq 0$ and $\delta_o^\alpha = 0$ define the ersatz material properties. 

\subsection{Internal contact}
In topology optimization, it has recently been suggested to employ the so-called third medium contact method (\citet{wriggers2013finite}) to model internal contact, cf. \citet{bluhm2021internal}, \citet{Andreas2023topology}, and \citet{dalklint2023computational}. In this work, the prospect of the third medium contact method is further extended to the problem of heat transfer between bodies in contact, such that the third medium now transfers heat when contact occurs, but otherwise acts as a thermally isolating material. Analogously to the purely mechanical problem, it can be shown that this behavior can be obtained exclusively via appropriate constitutive modeling of the third medium. 
This effect is readily seen by examining the definition of the heat flux in \eqrefK{heatFlux}. When the third medium is compressed, $J\rightarrow 0$ and since $\bm{F}^{-1}$ is inversely proportional to $J$, the heat flux in the normal direction of the surface in contact $\bm{n}\cdot\bm{q} \rightarrow \infty$.

In all real world thermal contact problems, a so called \emph{thermal contact resistance} exists due to irregularities in the contacting surfaces which impede the heat transfer. This contact resistance will give rise to a temperature jump across the contact interface at the macroscopic scale. Experiments show that the thermal contact resistance decreases with increased pressure, explained by the fact that the increased pressure causes more surface area to come into contact, increasing the heat transfer. Unfortunately, there exist no canonical form of the relationship between the thermal contact resistance, or its inverse; the thermal contact conductance, and the applied pressure. Indeed, this relationship is usually explored in an experimental setting, cf. e.g. \citet{rosochowska2003measurements} and \citet{semiatin1987determination}. However, there exist numerical models which approximates the thermal contact resistance. For example, \citet{aalilija2021simple} illustrated the use of a diffuse contact interface model to capture the thermal contact resistance. As we will show in the subsequent numerical examples, the third medium contact method also naturally captures the effect of this thermal contact resistance via the hyperelastic modeling of the third medium, i.e. the choice of the strain energy density function and material parameters.

In this work, traditional finite elements are employed as basis for the numerical solution of the boundary value problems from Eqs. \eqref{weak1} and \eqref{Helmholtz}. Unfortunately, this numerical analysis is plagued by excessively distorted finite elements in the third medium, which hinders the convergence. To improve the deformed element quality, we follow \citet{bluhm2021internal} and augment the strain energy density function \eqrefK{Psi} with a term that penalizes higher order strain components, i.e.
\begin{equation}
\Psi(\bm{u},\theta) \leftarrow \Psi(\bm{u},\theta) + \frac{k_r}{2}\mathbb{H}\bm{u}\, \tripledot \, \mathbb{H}\bm{u},
\label{PsiH}
\end{equation}
where $\mathbb{H}$ is the Hessian operator. Following \citet{bluhm2021internal}; $k_r = \bar{k}_r H^2 K_o$, where $\bar{k}_r$ is a small numerical parameter and $H$ is the characteristic height of the domain $\Omega$.

\subsection{Numerical solution procedure}
The fully coupled thermo-mechanical system \eqrefK{allR} is solved using Newton's method, i.e. in each step the full system \eqrefK{allR} is linearized to obtain
\begin{equation}
\ds\mathcal{{R}}(\bm{a} + d\bm{a};\delta\bm{a}) \approx \ds\mathcal{{R}}(\bm{a};\delta\bm{a}) + \mathcal{{R}}_{,a}(\bm{a};\delta\bm{a},d\bm{a}) = {0},
\label{dRu}
\end{equation} 
which is iteratively solved for $d\bm{a}$, such that $\bm{a} \leftarrow \bm{a} + d\bm{a}$. Without loss of generality, we limit the numerical examples to plane strain two dimensional problems. The presented thermo-mechanical topology optimization framework is implemented in Python where all tangent matrices and residual vectors are computed via the finite element library GetFEM (\citet{renard2020getfem}), which performs all underlying computations in C++. Design updates are based on a Python implementation (\citet{deetman2021mma}) of the Method of Moving Asymptotes (MMA) (\citet{svanberg1987method,svanberg2007mma}).

\subsection{Thermal contact}
\begin{figure}[H]
\centering
\includegraphics[width=0.75\textwidth]{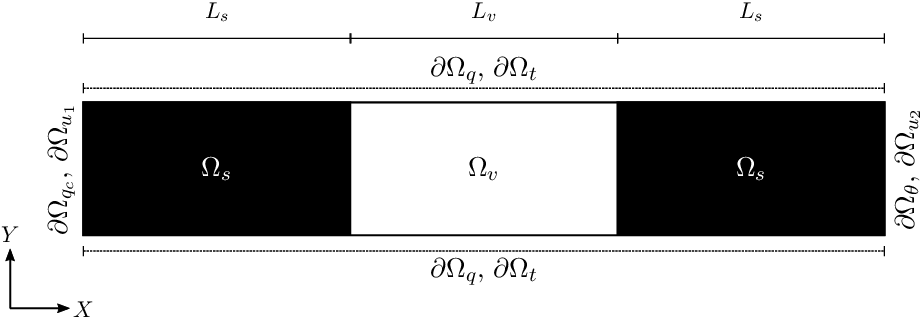}
\caption{The rod with dimensions $(L,H)$ = ($6$ cm $\times$ $1$ cm).}
\label{fig:rod}
\end{figure}

To illustrate how the third medium contact method facilitates thermal contact, the simple rod problem depicted in \figref{fig:rod} is considered. The rod is split into three parts of lengths $L^s =L^v = L/3$, such that two solid parts $\Omega_s$ with $E= 1$ MPa, $\nu=0.4$, and $\kappa = 10$ W/mK are separated by a third medium region $\Omega_v$ with $E_v = \delta_o^E E = 10^{-6} E$ and $\kappa_v = \delta_o^\kappa\kappa$, cf. \figref{fig:rod}. We investigate different values of $\delta_o^\kappa$ in the following analysis. For simplicity we assign $\alpha=0$ and focus only on the mechanical effect on the thermal problem, cf. \eqrefK{Psi}. We set $h = 1$ kW/m$^2$K and $\theta_\infty=20$ $\degree$C. Note that that $\partial\Omega_{u}$ is decomposed into the two complementary sets $ \partial\Omega_{u_1}$ and $\partial\Omega_{u_2}$.

To obtain self-contact in the rod, the Dirichlet condition $\bm{u} = (\bar{u},0)$ is applied on $\partial\Omega_{{u}_2}$, where $\bar{u}=-0.5L$, cf. \eqrefK{weak1}. On $\partial\Omega_{{u}_1}$, the pointwise value of $u_x$ and the average of $u_y$ are set to zero,. Also, we assign $\bar{\theta} = 200$ $\degree$C. The rod is discretized by $120\times 20$ quadratic quadrilateral finite elements, and nine point Gauss-Lobatto integration is used for improved stability in the third medium (\citet{faltus2024third}). The displacement $\bm{u}$ and temperature $\theta$ fields are discretized by incomplete Q2 elements, whereas the design variable $z$ and filtered $\zeta$ fields are discretized by Q0 and Q1 elements, respectively.   
The numerical solution is compared to an analytical solution derived from the 1D approximation of the thermal contact rod problem, which is obtained by solving the 1D heat equation
\begin{equation}
-\frac{d}{dX}\left(\kappa\frac{d\theta(X)}{dX}\right) = 0, \quad X\in\Omega,
\label{heat1}
\end{equation}
over the deformed rod after contact has been initiated, i.e. $\Omega=[0,L_c]$ where $L_c(\bar{u}) = L + \bar{u}$ is the deformed length of the rod. Equivalently to the 2D problem, Newton convection is present on the left hand side, i.e. $q(0) = h\left(\theta_\infty-\theta(0)\right)$, and a Dirichlet conditions is applied to the right hand side, i.e. $\theta(L_c) = \bar{\theta}$. Additionally, the thermal contact resistance is accommodated via the temperature jump $[[\theta(X=L_c/2)]] = -R_{th} q(L_c/2)$ at $X=L_c/2$, where $R_{th}$ is the thermal contact resistance and $q = -\kappa\frac{d\theta}{dX}$ is the 1D heat flux. We solve \eqrefK{heat1} over $\Omega$ by integrating twice over the two separate regions $0\leq X\leq L_c/2$ and $L_c/2< X\leq L_c$, and utilize the boundary conditions in combination with the aforementioned jump condition to couple the domain solutions. The final 1D temperature distribution becomes
\begin{equation}
\theta(X) = \begin{cases}
-\frac{1}{\kappa}(AX + B_1) \ \ \text{for} \ \ 0\leq X\leq L_c/2, \\[10pt]
-\frac{1}{\kappa}(AX + B_2) \ \ \text{for} \ \ L_c/2< X\leq L_c,
\end{cases}
\label{heat5}
\end{equation}
where $A = \frac{\theta_{\infty}-\bar{\theta}}{L_c/\kappa + R_{th} + 1/h}$, $B_1 = \kappa(\frac{A}{h} - \theta_{\infty})$ and $B_2 = -\kappa \bar{\theta}-AL_c$.

\begin{figure}[H]
\centering
\includegraphics[width=0.7\textwidth]{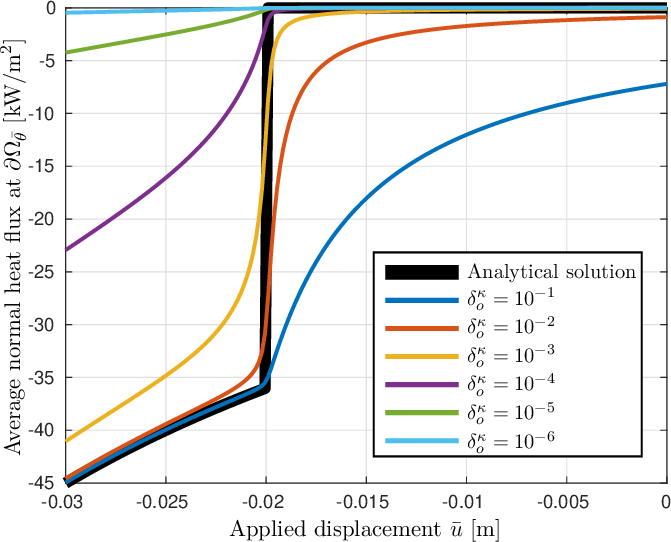}
\caption{The analytical and numerical average normal reaction fluxes versus the applied displacement for varying $\delta_o^\kappa$.} 
\label{fig:varyDeltakappa}
\end{figure}

First, a sharp contact interface between $\Omega_s$ and $\Omega_v$ is analyzed, i.e. the PDE filter is neglected. To investigate how the heat conductivity in the middle \say{void} region, $\kappa_v$, affects the heat transfer during contact, the average normal heat flux $\frac{1}{\vert\partial \Omega_{\theta}\vert}\int_{\Omega_{\theta}} \lambda_{\theta} \, dS$ is plotted versus the applied displacement $\bar{u}\in[0,-0.5L]$ for varying $\delta_o^\kappa$ in \figref{fig:varyDeltakappa}, where $\vert \partial\Omega_{\theta} \vert$ denotes the area. The results are compared to the analytical solution in \eqrefK{heat5} with perfect contact conditions, i.e. when $R_{th} = 0$, and different levels of deformation, i.e. deformed lengths $L_c$.
\noindent
It is noted that that a large premature heat transfer occurs for large values of $\delta_o^\kappa> 10^{-3}$, cf. \figref{fig:varyDeltakappa}. To reduce this premature heat transfer to acceptable levels, smaller values of $\delta_o^\kappa\leq 10^{-3}$ have to be considered. However, the simulations with $\delta_o^\kappa\leq 10^{-3}$ depicts a drastically reduced final heat flux magnitude in comparison to the theoretical maximum provided by the analytical solution. Fortunately, this phenomena proves useful in our analyzes, since it lets us simulate the aforementioned thermal contact resistance. Similarly to \citet{aalilija2021simple}, \figref{fig:varyDeltakappa} shows that the magnitude of the thermal contact resistance can be changed by tuning $\delta_o^\kappa$.

\begin{figure}[H]
\centering
\includegraphics[width=0.7\textwidth]{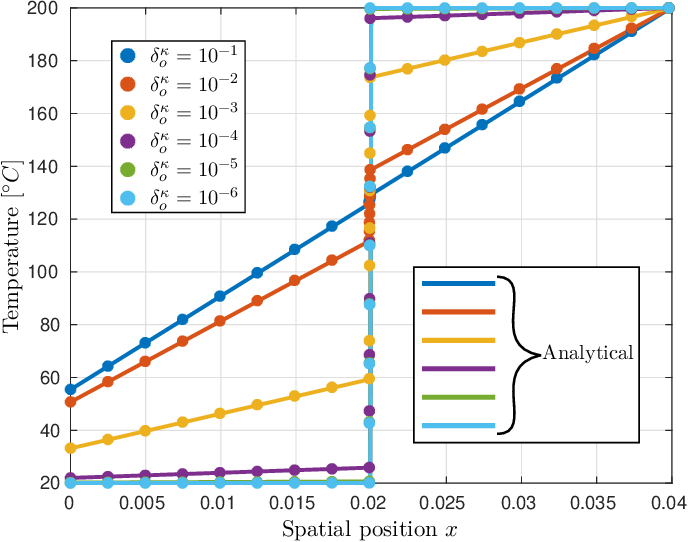}
\caption{The analytical and numerical temperature distributions over the $L_c\approx 0.04$ m rod for varying thermal contact resistances, i.e. varying $\delta_o^\kappa$.}
\label{fig:rodTempDist}
\end{figure}

\begin{figure}[H]
\centering
\includegraphics[width=0.7\textwidth]{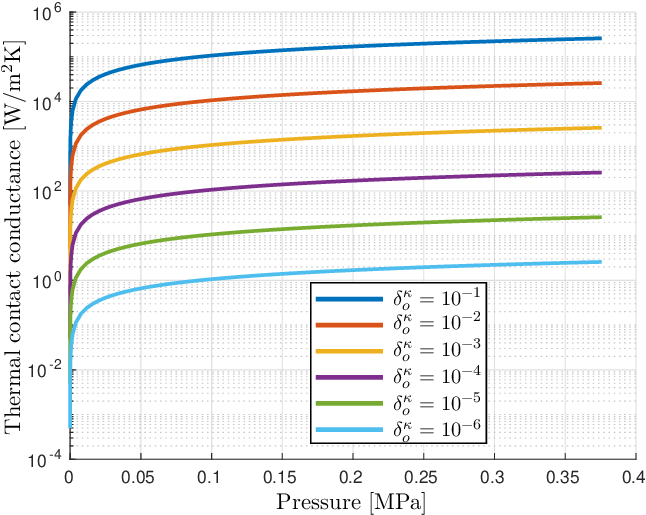}
\caption{The thermal contact conductance, $1/R_{th}$, versus the applied pressure for varying $\delta_o^\kappa$.}
\label{fig:conductanceVsPressure}
\end{figure}

Indeed, the 1D thermal contact resistance $R_{th}$ of our sharp interface third medium can be analytically approximated as
\begin{equation}
R_{th} = \frac{L_c^v}{\delta_o^\kappa\kappa},
\label{Rth}
\end{equation}
where $L_c^v$ is the deformed length of the third medium, i.e. void region. In \figref{fig:rodTempDist}, the analytical and numerical temperature distributions in the rod are compared for different thermal contact resistances, when $L_c\approx 0.4$ m and $R_{th}$ is obtained from \eqrefK{Rth} with $L_c^v$ computed from the numerical solution, cf. \eqrefK{heat5}. The results illustrate good correspondence between the analytical and numerical solutions.
Next, the implicit relationship between the thermal contact conductance and the applied pressure is investigated, cf. \eqrefK{Rth}. The obtained relationship is illustrated in \figref{fig:conductanceVsPressure} for varying $\delta_o^\kappa$.  

\begin{figure}[H]
\centering
\includegraphics[width=0.7\textwidth]{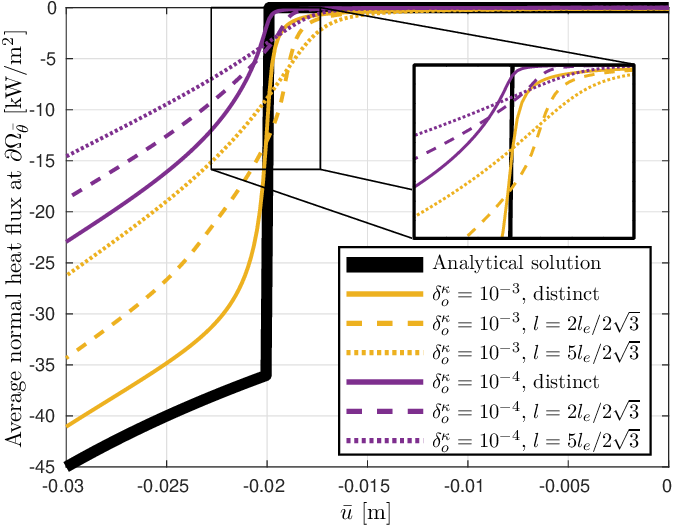}
\caption{The reaction flux versus the applied displacement for sharp and diffuse interfaces with varying filter radii and $\delta_o^\kappa$.}
\label{fig:rodPressure}
\end{figure}

\begin{figure}[H]
\centering
\includegraphics[width=\textwidth]{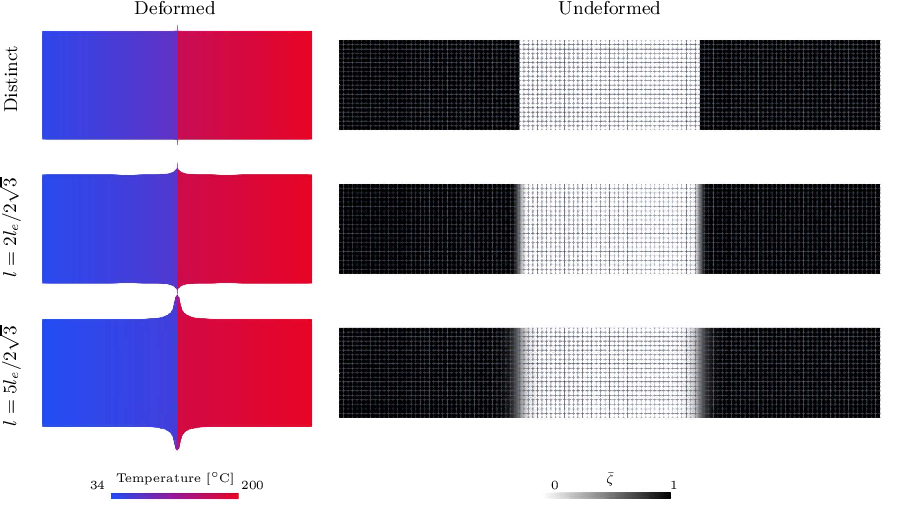}
\caption{The undeformed/deformed rods for the sharp/diffuse interface representations.}
\label{fig:rodPressureDef}
\end{figure}

The numerical investigation of the rod is concluded by considering a diffuse contact interface via the introduction of the PDE filter, cf. \eqrefK{Helmholtz}. In \figref{fig:rodPressure} the average normal heat flux on $\partial\Omega_{\theta}$ is plotted versus the applied displacement $\bar{u}$ for the sharp and diffuse interfaces with varying filter radii and $\delta_o^\kappa$. As expected, it is confirmed that the introduction of the diffuse interface increases the thermal contact resistance.
\noindent
Finally, the undeformed/deformed rods with sharp/diffuse interfaces for $\delta_o^\kappa = 10^{-4}$ are depicted in \figref{fig:rodPressureDef}. A visual inspection of the temperature distributions shown in the bottom row of \figref{fig:rodPressureDef} confirms the increased contact resistance for the rods with diffuse interface. It is noted that large transverse displacements occur at the contact interface when introducing the filter, which are believed to be caused by the reduced stiffness of the intermediate volume fraction regions (cf. the SIMP scheme in \eqrefK{matInt}).
Based on the above numerical study we conclude that; 1) the third medium contact method is capable of modeling contact with a non-vanishing thermal contact resistance, and 2) the magnitude of the thermal contact resistance can be controlled via the ersatz material parameters; more specifically $\delta_o^\kappa$, and the gap size; i.e. $L_c^v$ in the 1D model. In other words, $\delta_o^\kappa$ must not solely be viewed as a numerical parameter, but must be calibrated based on experiments. In this work, we nevertheless limit ourselves to numerical studies, wherefore $\delta_o^\kappa \in \{10^{-4},10^{-3}\}$ is used, without loss of generality. It is also concluded that the diffuse contact interfaces that appear due to the filter increase the thermal contact resistance. However, in our optimization, a continuation scheme for the thresholding (cf. \eqrefK{Heavi}) is used, wherefore we expect the final design to be approximately binary, i.e. the above 2) conclusion still holds.

\subsection{Optimization problem}
The optimization problems solved can be written in the following format
\begin{equation}
\mathbb{P}\ \begin{cases}
\begin{array}{ll}
\underset{z}{\text{min}} \ C({\zeta},\bm{a}), \\[8pt]
\text{s.t.} 
\begin{cases}
{\mathcal{R}}(\bm{a};\delta\bm{a}) = {0}, \\[5pt]
\mathcal{R}_\zeta(\zeta,z;\delta\zeta) = {0}, \\[5pt]
\ds g = \int_\Omega H_{\beta,\eta_d}(\zeta)\, dV - V^* \leq 0,\\[8pt]
0 \leq z \leq 1,
\end{cases}
\end{array}
\end{cases}
\label{optProb}
\end{equation}
where we define the objective function $C$ and an upper bound volume constraint $g$, ensuring lightweight designs. In this work, thin features are penalized in the designs via the inclusion of the volume constraint on the dilated design field $H_{\beta,\eta_d}(\zeta)$, where $\eta_d = 0.4$ (cf. \citet{Andreas2023topology}).

\subsubsection{Sensitivity analsysis}
The topology optimization problem is solved in a staggered fashion using the adjoint method, wherefore in each design iteration, \eqrefK{weak1} is solved exactly. In the adjoint method, the augmented objective function 
\begin{equation}
\begin{array}{ll}
\ds C^*({\zeta},\bm{a}) = &\ds C({\zeta},\bm{a}) + \mathcal{{R}}_a({\zeta},\bm{a};\bm{\mu}_a) + \mathcal{R}_\zeta(\zeta,z;\mu_\zeta),
\end{array}
\label{augmentedg}
\end{equation}
is defined, where $\bm{\mu}_a = (\bm{\mu}_u,\mu_\theta,\mu_{\lambda_u},\mu_{\lambda_\theta})$ is the adjoint vector.
Taking the first variation with respect to the design variable field $z$ renders
\begin{equation}
\begin{array}{ll}
\delta C^* =& \mathcal{R}_{\zeta,z}(\zeta,z;\mu_\zeta, \delta z) \\[10pt]
&+ C^*_{,\zeta} + \mathcal{{R}}_{,\zeta}({\zeta},\bm{a};\bm{\mu}_a,\delta \zeta) + \mathcal{R}_{\zeta,\zeta}(\zeta,z;\mu_\zeta, \delta \zeta) \\[10pt]
&+ C^*_{,a} + \mathcal{{R}}_{,a}({\zeta},\bm{a};\bm{\mu}_a,\delta \bm{a}),
\end{array}
\label{varg}
\end{equation}
and to annihilate the implicit variation $\delta\bm{a}$, $\bm{\mu}_a$ is assigned such that 
\begin{equation}
C^*_{,a} + \mathcal{{R}}_{,a}({\zeta},\bm{a};\bm{\mu}_a,\delta \bm{a}) = \bm{0}.
\label{adjoints}
\end{equation}
Next, the implicit variation $\delta\zeta$ is annihilated by assign $\mu_\zeta$ as the solution to
\begin{equation}
C^*_{,\zeta} + \mathcal{{R}}_{,\zeta}({\zeta},\bm{a};\bm{\mu}_a,\delta \zeta) + \mathcal{R}_{\zeta,\zeta}(\zeta,z;\mu_\zeta, \delta \zeta) = 0.
\label{adjointZeta}
\end{equation}
Finally, the sensitivity expression reduces to
\begin{equation}
\delta C^* =  \mathcal{R}_{\zeta,z}(\zeta,z;\mu_\zeta, \delta z).
\label{sens}
\end{equation}
The sensitivity of the volume constraint $g$ is trivial and therefore omitted in this presentation.

\section{Numerical examples}
In the following numerical examples, thermal regulators that actuate solely via thermal expansion/shrinking are designed. Therefore, we assign a non-zero thermal expansion coefficient $\alpha$ and only apply homogeneous Dirichlet conditions on $\partial\Omega_{u}$. The same finite element spaces as presented in the previous example is used to discretize the fields. The material parameters used in the subsequent examples are stated in Tab. \ref{tab:mat}.
\begin{table}[H]
        \centering
        \begin{tabular}{lr}
        \hline
            Young's Modulus $E$ [MPa] & $1$ \\
            Poisson's ratio $\nu$  & $0.4$ \\ 
            Thermal conductivity $\kappa$ [W/mK] & $10$\\ 
            Thermal expansion coefficient $\alpha$ [1/K] & $10^{-4}$\\ 
            Reference temperature $\theta_o$ [\degree C]  & $20$ \\
          \hline    
        \end{tabular}
        \caption{The material parameters.}
        \label{tab:mat}
\end{table}
\noindent
The Heavside threshold parameter $\beta$ is initially set to $\beta^{\text{ini}}$, and is after 300 design iterations increased $\beta\leftarrow 2\cdot\beta$ every 50th design iteration until $\beta^{\text{max}}$ is reached. Optimization and mesh parameters that are common to all subsequent examples are presented in Tab. \ref{tab:opt}.
\begin{table}[H]
        \centering
        \begin{tabular}{lr}
        \hline
        	Maximum volume $V^*$ & $0.4\vert\Omega\vert$\\
        	Mesh size & $160 \times 80$ \\ 
        	Element side length $l_e$ [m] & 0.0625\\ 
        	Filter radius $l$ [m] & $2l_e/2\sqrt{3}$ \\
            Ersatz material scaling $\delta_o^E$ & $10^{-6}$ \\
            Void regularization weight $\bar{k}_r$ & $10^{-6}$\\
            Penalty parameter $p$ & 3\\
          \hline    
        \end{tabular}
        \caption{The optimization and mesh parameters.}
        \label{tab:opt}
\end{table}

\subsection{Thermal switch}
\begin{figure}[H]
\centering
\includegraphics[width=0.5\textwidth]{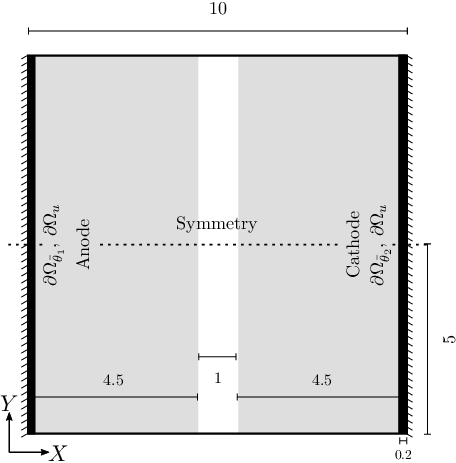}
\caption{The first design domain. The design variable field $z$ is prescribed $z=1$ in the black regions throughout the optimization. The gray regions are initially assigned $z = 0.3$, whereas the white region is $z=10^{-4}$. All dimensions are in millimeters.}
\label{fig:switch}
\end{figure}

\begin{figure}[H]
	\centering
	\includegraphics[width=\textwidth]{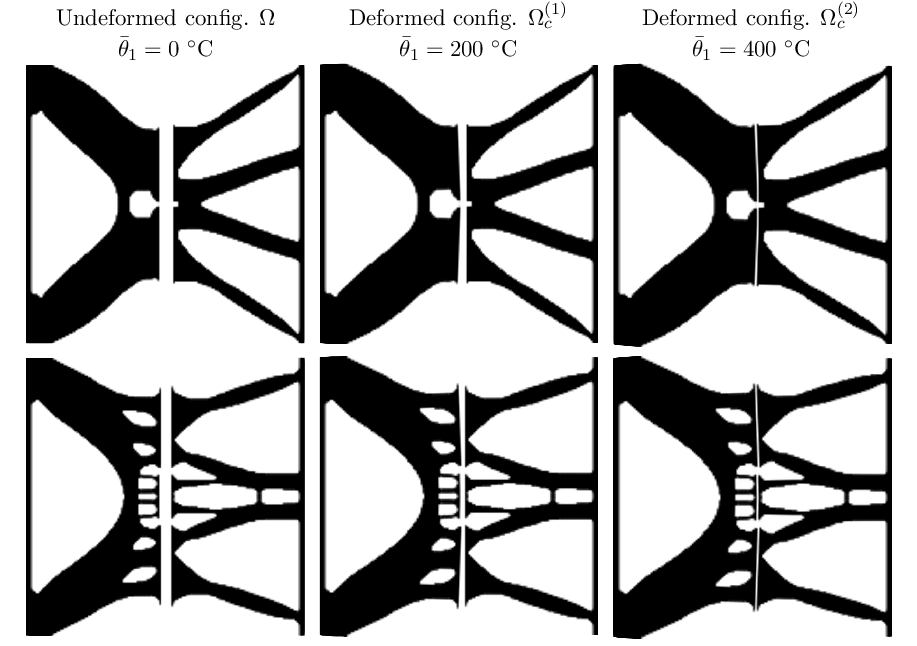}
	\caption{The thermal switch designs in their undeformed and deformed configurations for $\delta_o^\kappa=10^{-3}$ (top row) and $\delta_o^\kappa=10^{-4}$ (bottom row).}
	\label{fig:switch_diffK}
\end{figure}

\begin{figure}[H]
	\centering
	\includegraphics[width=0.8\textwidth]{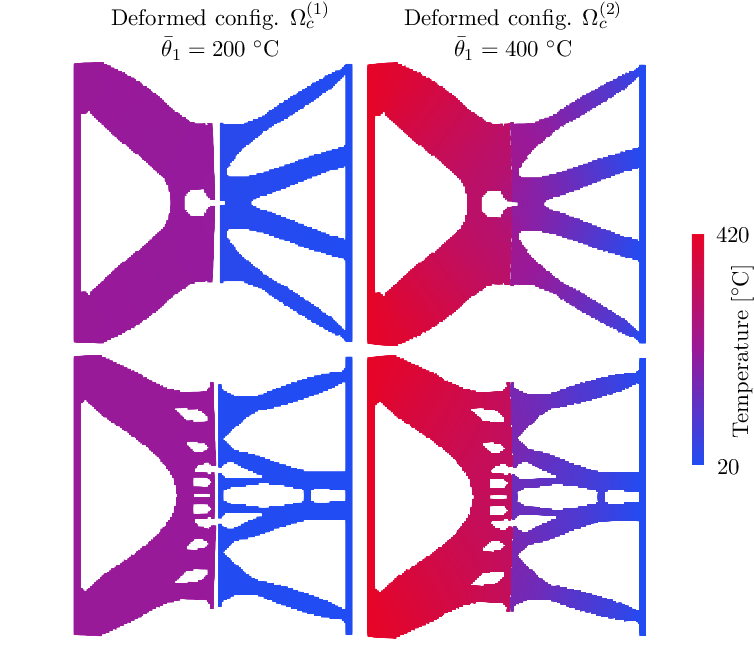}
	\caption{The temperature distributions over the \figref{fig:switch_diffK} designs for $\delta_o^\kappa=10^{-3}$ (top row) and $\delta_o^\kappa=10^{-4}$ (bottom row). }
	\label{fig:switch_temp_diffK}
\end{figure}

\begin{figure}[H]
	\centering
	\includegraphics[width=\textwidth]{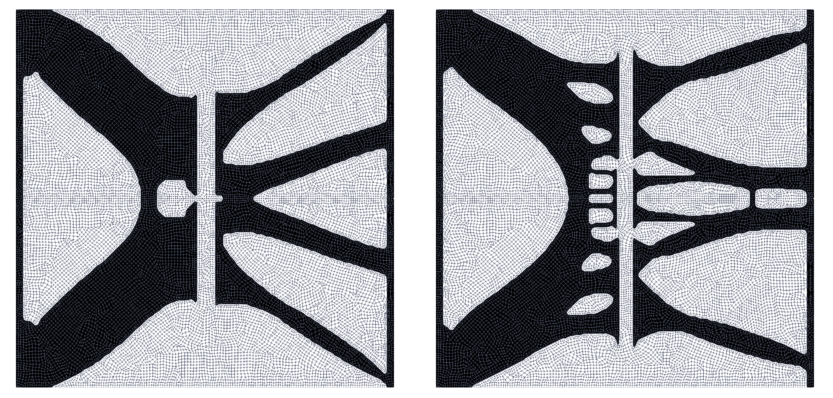}
	\caption{The body-fitted meshes of the \figref{fig:switch_diffK} designs with $\delta_o^\kappa = 10^{-3}$ (left) and $\delta_o^\kappa = 10^{-4}$ (right).}
	\label{fig:switch_fitted_mesh}
\end{figure}

\begin{figure}[H]
	\centering
	\includegraphics[width=0.7\textwidth]{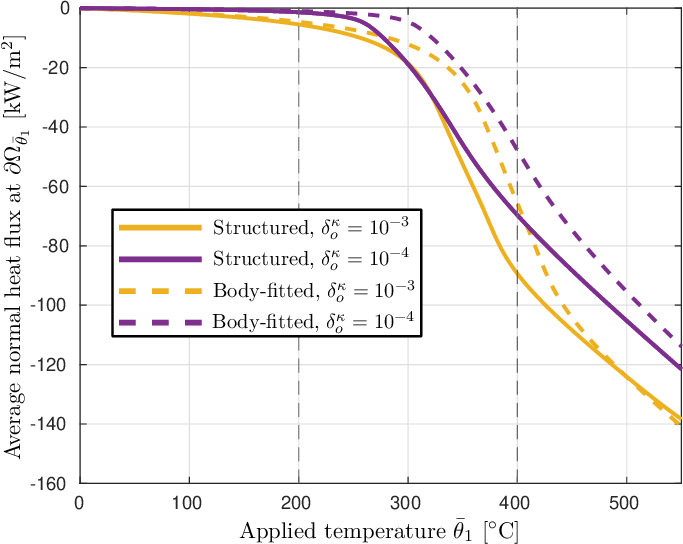}
	\caption{The average normal heat flux versus the applied temperature for the \figref{fig:switch_diffK} designs discretized on structured and body-fitted meshes. The dashed black lines highlights the target values $\bar{\theta}_1\in\{200, 400\}$ $\degree$C.}
	\label{fig:switch_fitted}
\end{figure}

\begin{figure}[H]
	\centering
	\includegraphics[width=0.8\textwidth]{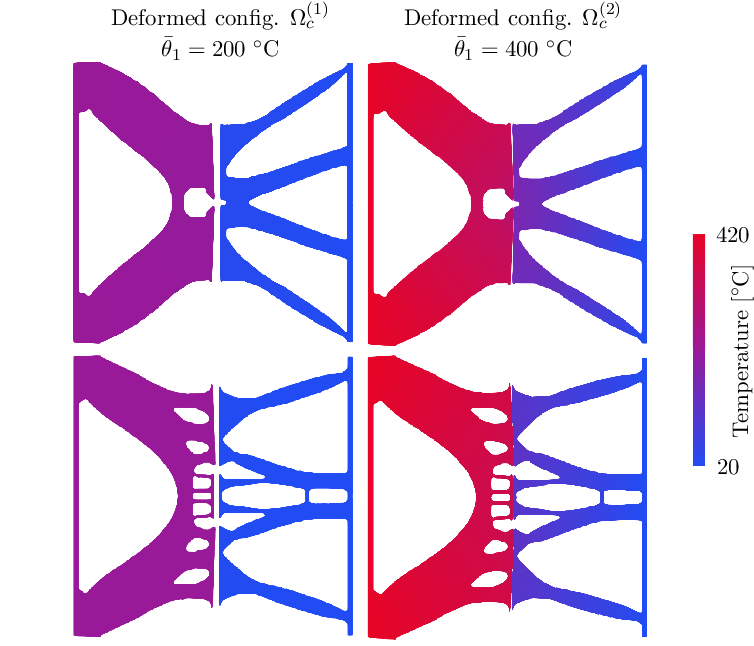}
	\caption{The temperature distributions over the \figref{fig:switch_fitted_mesh} designs for $\delta_o^\kappa = 10^{-3}$ (top row) and $\delta_o^\kappa = 10^{-4}$ (bottom row) in their deformed configurations.}
	\label{fig:switch_fitted_temp}
\end{figure}

In the first example, contact-aided thermal switches are designed. The design domain contains an anode ($\partial\Omega_{\bar{\theta}_1}$) and a cathode ($\partial\Omega_{\bar{\theta}_2}$), at which the temperature is controlled, cf. \figref{fig:switch}. The temperature at the anode is varied, whereas the temperature at the cathode is kept constant. To mimic the behavior of a thermal switch, the heat transfer between the anode and the cathode should vanish until a certain critical temperature difference is reached, after which the heat transfer should be facilitated. All boundaries without specific labels are part of $\partial\Omega_q$ and $\partial\Omega_t$, i.e. are isolated and traction free. Note that $\partial\Omega_{q_c} = \emptyset$ and that $\partial\Omega_{\theta}$ is decomposed into the two complementary sets $ \partial\Omega_{{\theta}_1}$ and $\partial\Omega_{{\theta}_2}$ over which $\bar{\theta}_1$ and $\bar{\theta}_2$ are prescribed, respectively. As depicted in \figref{fig:switch}, a non-uniform volume fraction distribution with a centered gap is utilized as initial guess. By moving the initial position of this gap, similar albeit shifted designs are obtained, which indicates that we are dealing with a highly non-convex optimization problem. It has been found that a uniform initial design can also produce meaningful designs; however, the convergence of the optimization process is significantly hindered.
\noindent
To design thermal switches, the objective function
\begin{equation}
C = \int_{\partial\Omega_{\bar{\theta}_1}}\left(-w_{1} \lambda_{\theta}^{(1)} + w_{2} \lambda_{\theta}^{(2)} \right) \, dS, 
\label{objSwitch}
\end{equation}
is employed, where $(\,\cdot \,)^{(i)}$ denotes quantities evaluated at target point $i = 1,2$ and $w_{i}$ are weights. Since the direction of the temperature gradient is known, the above objective function promotes vanishing heat flux at the first target point ($i=1$) and a large (negative) heat flux at the second target point ($i=2$). The parameters for the switch example are summarized in Tab. \ref{tab:diode}.
\begin{table}[H]
        \centering
        \begin{tabular}{lr}
        \hline
        Threshold parameters $(\beta^{\text{ini}},\beta^{\text{max}})$ &  $(2,16)$\\
        Target temperatures $(\bar{\theta}_1^{(1)},\bar{\theta}_2^{(1)})$ [$\degree $C] & $(200, 0)$  \\ 
        Target temperatures $(\bar{\theta}_1^{(2)},\bar{\theta}_2^{(2)})$ [$\degree $C] & $(400, 0)$ \\ 
        Weights $(w_{1},w_{2})$ & $(2\cdot10^3, 10^3)$\\
        \hline
        \end{tabular}
        \caption{The numerical parameters for the switch example, unless stated otherwise. Note that only half the domain in \figref{fig:switch} is discretized due to symmetry.}
        \label{tab:switch}
\end{table}
In \figref{fig:switch_diffK}, two optimized thermal switch designs with different thermal contact resistances, i.e. $\delta_o^\kappa$,  are depicted. A glance at the designs in their undeformed configurations reveals that the $\delta_i^\kappa = 10^{-4}$ design has a smaller initial gap between the contact surfaces than the $\delta_i^\kappa = 10^{-3}$ design. This is expected since the thermal contact resistance of the $\delta_i^\kappa = 10^{-4}$ design is larger, i.e. it isolates better at the first target point, and must exert a greater contact pressure than the $\delta_i^\kappa = 10^{-3}$ design to reach the same heat flux at the second target point.
\noindent
In \figref{fig:switch_temp_diffK} the temperature distributions in the deformed configurations are shown for the \figref{fig:switch_diffK} designs. Again, the effect of the thermal contact resistances is noted. Indeed, the $\delta_o^\kappa = 10^{-3}$ design exhibits a smoother temperature gradient across the contact interface when contact occurs, compared to the $\delta_o^\kappa = 10^{-4}$ design.
The boundaries of the \figref{fig:switch_diffK} designs are poorly resolved, which can adversely affect the performance of the realized designs. Therefore, additional post-processing analyzes are conducted, in which the performance of the structured mesh is compared to body-fitted mesh representations. The body-fitted meshes in \figref{fig:switch_fitted_mesh} have been generated using GMSH (\citet{geuzaine2009gmsh}), based on the $\zeta \approx 0.5$ contour line. The fitted meshes consist of Q2 and P2 elements which are integrated using nine and seven Gauss-points, respectively. Again, for ease of implementation, the third medium contact method is used to model the thermo-mechanical contact, wherefore both the solid and void regions are meshed. 
In \figref{fig:switch_fitted}, the average normal heat flux versus the applied temperature is plotted for the \figref{fig:switch_diffK} designs discretized on structured and body-fitted meshes. As expected, the $\delta_o^\kappa=10^{-4}$ design exhibits less heat flux at both target points compared to the $\delta_o^\kappa=10^{-3}$ design. Seemingly, there exists a trade-off between good isolating properties (smaller $\delta_o^\kappa$), and good heat transfer properties (larger $\delta_o^\kappa$). It is also noted that the designs exhibit premature heat transfer, i.e. contact, on the structured mesh, in relation to the body-fitted mesh. This is in line with the results from the rod problem when using discrete or diffuse contact interfaces, cf. \figref{fig:rodPressure}. Indeed, it is noted that after a critical temperature of $\bar{\theta} \approx 500$ $\degree$C is reached, the heat flux of the $\delta_o^\kappa = 10^{-3}$ body-fitted design exceeds its structured mesh counterpart, similar to the observed behavior of the rod problem. The importance of the interface resolution was also scrutinized in \citet{bluhm2021internal} and \citet{dalklint2023computational}, who came to similar conclusions. The $\delta_o^\kappa = 10^{-4}$ design doesn't exhibit the same behavior at the applied temperature range due to the presence of a softening instability mode, leading to bending of the right hand side top and bottom structural members, effectively reducing the contact pressure and thereby heat flux. To avoid instabilities at operating temperature conditions we could follow \citet{Andreas2023topology} and penalize small tangent stiffness values, or introduce buckling constraints via a linearized buckling analysis, but this is deemed outside the scope of this paper. 
\noindent
In \figref{fig:switch_fitted_temp} the temperature distributions over the deformed designs discretized by body-fitted meshes are depicted. Qualitatively, the behavior of \figref{fig:switch_temp_diffK} is confirmed, i.e. the $\delta_o^\kappa=10^{-3}$ design exhibits a smoother temperature gradient across the contact interface than the $\delta_o^\kappa=10^{-4}$ design, as expected.
\noindent
Based on this post-processing analysis, is is confirmed that there exist some discrepancies between the analyzed and realized design. However, their extent is limited in the sense that the general behavior of the design remains. Therefore, for the following design examples, we refrain from conducting the similar post-processing analysis.

\subsection{Thermal diode}
\begin{figure}[H]
	\centering
	\includegraphics[width=\textwidth]{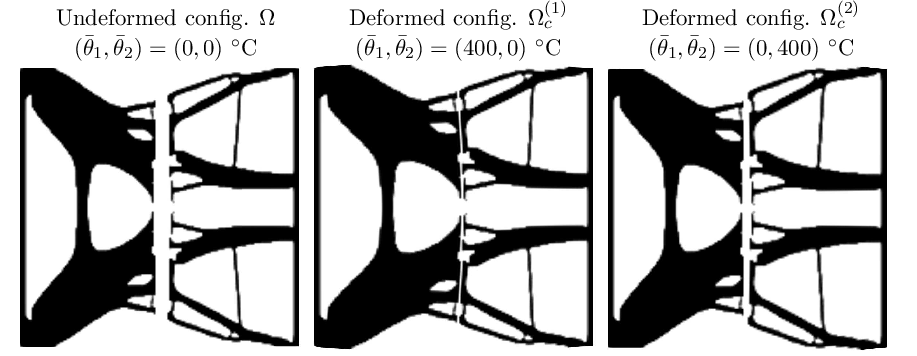}
	\caption{The thermal diode design in its undeformed and deformed configurations.}
	\label{fig:diode_des}
\end{figure}

\begin{figure}[H]
	\centering
	\includegraphics[width=0.8\textwidth]{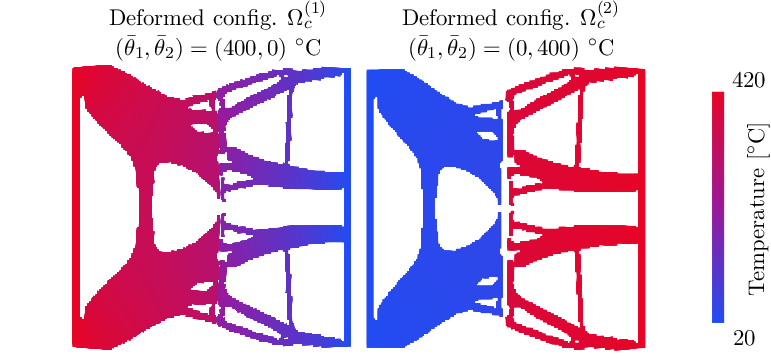}
	\caption{The temperature distributions over the \figref{fig:diode_des} design in its deformed configurations. }
	\label{fig:diode_temp}
\end{figure}

\begin{figure}[H]
	\centering
	\includegraphics[width=0.7\textwidth]{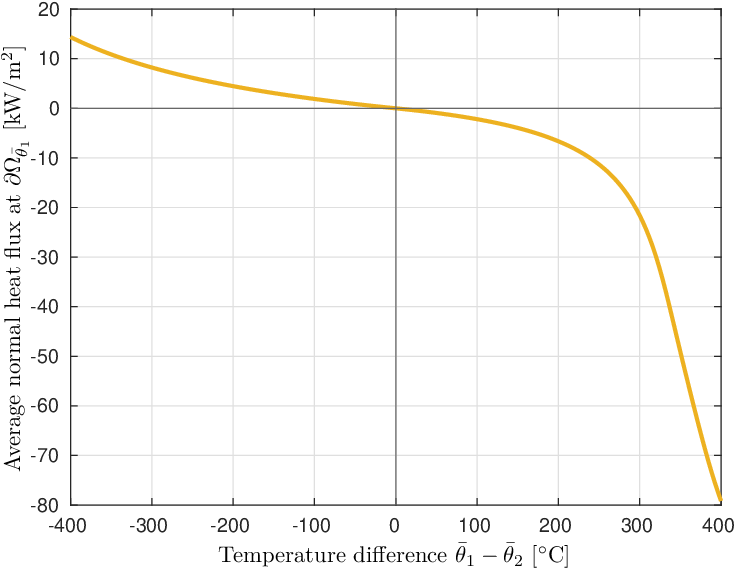}
	\caption{The average normal heat flux versus the difference in temperature on $\partial\Omega_{\bar{\theta}_1}$ and $\partial\Omega_{\bar{\theta}_2}$ for the \figref{fig:diode_des} design.}
	\label{fig:diode_plot}
\end{figure}

In the second example, a contact-aided thermal diode is designed. Again, the design domain is that of \figref{fig:switch}, but now two \say{load} cycles are performed: 1) $\partial\Omega_{\bar{\theta}_1}$ is heated whereas the temperature at $\partial\Omega_{\bar{\theta}_2}$ constant, 2) the polarity is switched such that $\partial\Omega_{\bar{\theta}_1}$ has constant temperature and $\partial\Omega_{\bar{\theta}_2}$ is heated. Our diode should rectify the heat \say{current}, such that the heat transfer between the terminals is facilitated during load cycle 1), whereas it is impeded during load cycle 2). To this end, the objective function
\begin{equation} 
C = w_{1}\int_{\partial\Omega_{\bar{\theta}_1}} \left(\lambda_{\theta}^{(1)} + \lambda_{\theta}^{(2)} \right) \, dS, 
\label{objDiode}
\end{equation}
is chosen, which promotes a large (negative) heat flux at the first target point ($i=1$), i.e. during load cycle 1), and a vanishing heat flux at the second target point ($i=2$), i.e. during load \mbox{cycle 2)}. The parameters for the diode example are summarized in Tab. \ref{tab:diode}.
\begin{table}[H]
        \centering
        \begin{tabular}{lr}
        \hline
        Threshold parameters $(\beta^{\text{ini}},\beta^{\text{max}})$ &  $(2,16)$\\
        Ersatz material scaling $\delta_o^\kappa$ & $10^{-3}$ \\
        Target temperatures $(\bar{\theta}_1^{(1)},\bar{\theta}_2^{(1)})$ [$\degree $C] & $(400,0)$  \\ 
        Target temperatures $(\bar{\theta}_1^{(2)},\bar{\theta}_2^{(2)})$ [$\degree $C] & $(0,400)$ \\ 
        Weight $w_{1}$ & $10^3$\\
        \hline
        \end{tabular}
        \caption{The numerical parameters for the diode example. Note that only half the domain in \figref{fig:switch} is discretized due to symmetry.}
        \label{tab:diode}
\end{table}
An optimized thermal diode design is depicted in \figref{fig:diode_des}. As envisioned, the thermal expansion renders a contact connection between the terminals during the load cycle 1), which facilitates the heat transfer. On the other hand, the design avoids contact during load cycle 2) to minimize the heat transfer.
\noindent
The temperature distributions in \figref{fig:diode_temp} confirm a larger heat transfer between the terminals in load cycle 1), compared to load cycle 2).
\noindent
This behavior is again confirmed in \figref{fig:diode_plot}, where the average normal heat flux on $\partial\Omega_{\bar{\theta}_1}$ is plotted versus the temperature difference between $\partial\Omega_{\bar{\theta}_1}$ and $\partial\Omega_{\bar{\theta}_2}$. Clearly, our diode design exhibits larger  heat transfer when the temperature difference is positive, than when negative.

\subsection{Thermal triode}
\begin{figure}[H]
\centering
\includegraphics[width=0.6\textwidth]{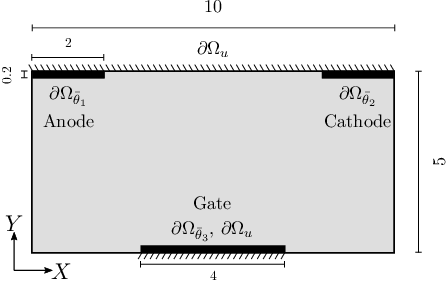}
\caption{The second design domain. The design variable field $z$ is prescribed $z=1$ in the black regions throughout the optimization. The gray region is initially assigned $z = 0.3$. All dimensions are in millimeters.}
\label{fig:triode}
\end{figure}

\begin{figure}[H]
	\centering
	\includegraphics[width=\textwidth]{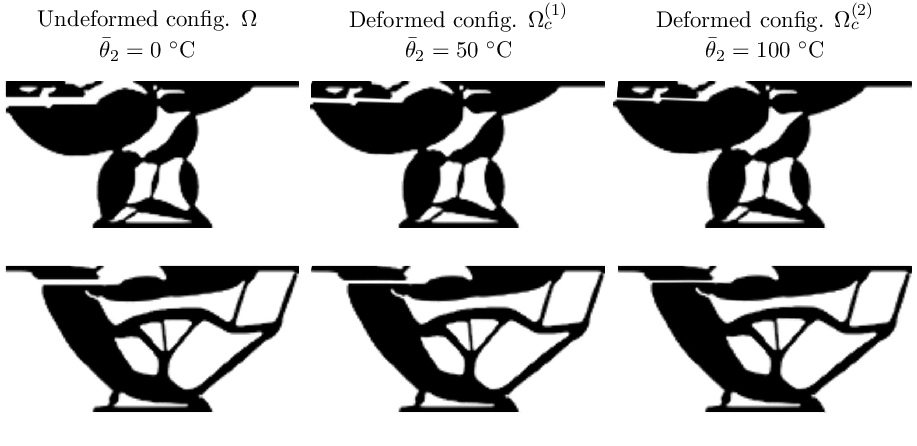}
	\caption{The thermal triode designs in their undeformed and deformed configurations.}
	\label{fig:triode_des}
\end{figure}

\begin{figure}[H]
	\centering
	\includegraphics[width=0.8\textwidth]{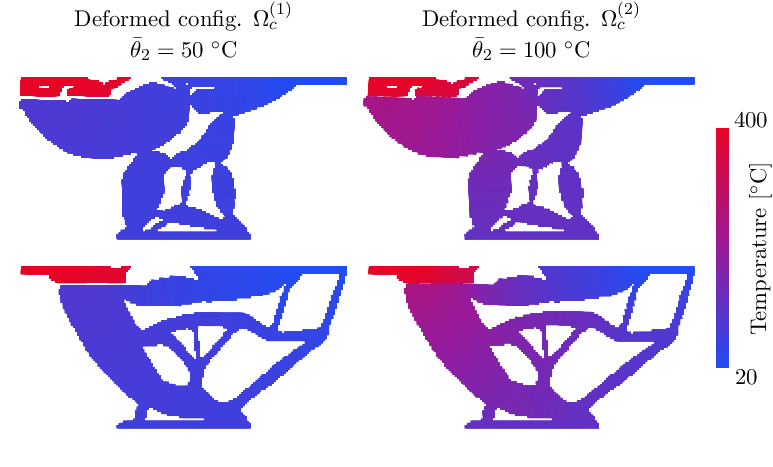}
	\caption{The temperature distributions over the \figref{fig:triode_des} designs in their deformed configurations. }
	\label{fig:triode_temp}
\end{figure}

\begin{figure}[H]
	\centering
	\includegraphics[width=0.7\textwidth]{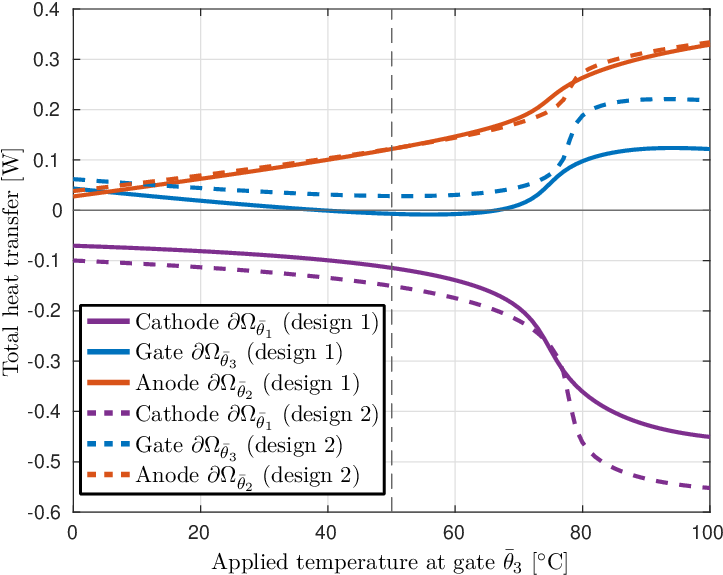}
	\caption{The total heat transfer at the anode, gate and cathode versus the applied temperature for the \figref{fig:triode_des} designs. The dashed black line highlights the target value $\bar{\theta}_1 = 50$ $\degree$C. }
	\label{fig:triode_plot}
\end{figure}

In the last example, we design contact-aided thermal triodes. This design domain comprises three terminals: one anode ($\partial\Omega_{\bar{\theta}_1}$), one cathode ($\partial\Omega_{\bar{\theta}_2}$) and one gate ($\partial\Omega_{\bar{\theta}_3}$), cf. \figref{fig:triode}. There exists a constant large temperature difference between the anode and the cathode, but the variable temperature at the gate dictates whether heat transfer between the anode and cathode is facilitated or impeded. Again, $\partial\Omega_{q_c} = \emptyset$ and all boundaries without specific labels are part of $\partial\Omega_q$ and $\partial\Omega_t$, i.e. are isolated and traction free. In this example, $\partial\Omega_{\theta}$ is decomposed into the three complementary sets $ \partial\Omega_{\bar{\theta}_1}$, $ \partial\Omega_{\bar{\theta}_2}$ and $\partial\Omega_{\bar{\theta}_3}$ over which $\bar{\theta}_1$, $\bar{\theta}_2$ and $\bar{\theta}_3$ are prescribed, respectively. In this example, a uniform initial volume fraction distribution is utilized.
\noindent
To design the thermal triodes, the objective function
\begin{equation}
\begin{array}{ll}
\ds C = &\ds \frac{w_{1}}{\vert\partial\Omega_{\bar{\theta}_1} \vert}\int_{\partial\Omega_{\bar{\theta}_1}}\left(- \lambda_{\theta}^{(1)} + \lambda_{\theta}^{(2)} \right) \, dS+ \frac{w_{2}}{\vert\partial\Omega_{\bar{\theta}_3} \vert}\int_{\partial\Omega_{\bar{\theta}_3}}\left( \left(\lambda_{\theta}^{(1)}\right)^2 + \left(\lambda_{\theta}^{(2)}\right)^2 \right) \, dS,
\end{array}
\label{objTriode}
\end{equation}
is chosen. This objective function promotes vanishing heat transfer at the first target point ($i=1$), and a large heat transfer between the anode and the cathode at the second target point ($i=2$), whilst also promoting a vanishing heat flux at the gate. The minus sign is introduced in the first term of \eqrefK{objTriode} since the heat flux in the normal direction will be negative at the anode, whereas the third and fourth terms of \eqrefK{objTriode} are squared since the sign of the normal heat flux at the gate is unknown. Using similar arguments, a positive sign is assigned to the second term of \eqrefK{objTriode}. The influence of the penalty of the average normal heat flux at the gate is investigated by designing two triodes. The first design (design 1) is obtained using a weight $w_2>0$, whereas the second design (design 2) is obtained when $w_2=0$. The parameters for the triode examples are summarized in Tab. \ref{tab:triode}.
\begin{table}[H]
        \centering
        \begin{tabular}{lr}
        \hline
        Threshold parameters $(\beta^{\text{ini}},\beta^{\text{max}})$ &  $(2,8)$\\
        Ersatz material scaling $\delta_o^\kappa$ & $10^{-3}$ \\
        Target temperatures $(\bar{\theta}_1^{(1)},\bar{\theta}_2^{(1)},\bar{\theta}_3^{(1)})$ [$\degree $C] & $(380, 50, 0)$  \\ 
        Target temperatures $(\bar{\theta}_1^{(2)},\bar{\theta}_2^{(2)},\bar{\theta}_3^{(2)})$ [$\degree $C] & $(380, 100, 0)$ \\ 
        Weights design 1 $(w_1,w_2)$ & $(10^3,10^3)$\\
        Weights design 2 $(w_1,w_2)$ & $(10^3,0)$\\
        \hline
        \end{tabular}
        \caption{The numerical parameters for the triode examples.}
        \label{tab:triode}
\end{table}
The thermal triode designs are depicted in \figref{fig:triode_des} in their undeformed and deformed configurations. It is seen that as the temperature at the gate is increased, the thermal expansion renders a connection between the anode and the cathode via structural contact, such that the heat transfer is maximized. It is noted that the gap to the anode is smaller in \mbox{design 2}, compared to \mbox{design 1}. In \figref{fig:triode_temp}, the temperature distributions are depicted in the deformed configurations. 
\noindent
Finally, the total heat transfer $Q_j = \int_{\Omega_{\bar{\theta}_j}} \lambda_{\theta} \, dS$, at the anode ($j=1$), gate ($j=3$) and cathode ($j=2$) are plotted versus the applied temperature in \figref{fig:triode_plot}. It is confirmed that the optimization renders a design that can amplify the control signal, i.e. the absolute value of the total heat transfer at the gate is smaller than that of the cathode. We conclude that the penalty to the average normal heat flux at the gate increases the efficiency of the triode, in the sense that less heat is lost through the gate for \mbox{design 1} than for \mbox{design 2}. It is also noted that the total heat transfer at the gate changes sign during actuation, and that there are two points ($\bar{\theta}_3 \approx 38$ $\degree$C and $\bar{\theta}_3 \approx 70$ $\degree$C) at which the total heat transfer vanishes at the gate. It is emphasized that this does not imply that the pointwise normal heat flux at the gate vanishes. 


\section{Conclusions}
In this work, we have via several numerical examples exemplified how the third medium contact method can be utilized to model complex thermo-mechanical contact problems. Indeed, it has been shown that the method inherently captures the thermal contact resistance that is present in real-world contact problems via the third medium, i.e. ersatz material properties. It has specifically been shown that the magnitude of the thermal contact resistance can be tuned via the ersatz material thermal conductivity contrast, $\delta_o^\kappa$. The implemented thermo-mechanical contact framework was shown to correspond well to an analytical solution.

The fully coupled thermo-mechanical numerical framework was applied to solve several topology optimization problems involving the design of thermal regulators that leverage contact as a non-linear mechanism. More specifically, we designed thermal switches, triodes and a diode. To validate the designs, a post-processing analysis using body-fitted meshes was conducted. 

\section*{Acknowledgments}
This work was sponsored by Independent Research Fond Denmark via the TopCon Project (case number 1032- 00228B). 

\bibliographystyle{elsarticle-harv}
\bibliography{thermalContact}

\end{document}